# Unified Quantification of Quantum Defects in Small-Diameter Single-Walled Carbon Nanotubes by Raman Spectroscopy

*Finn L. Sebastian[1], Felicitas Becker[1], Yohei Yomogida[2], Yuuya Hosokawa[2], Simon Settele[1], Sebastian Lindenthal[1], Kazuhiro Yanagi[2], Jana Zaumseil[1]\**

[1]Institute for Physical Chemistry, Universität Heidelberg, D-69120 Heidelberg, Germany

[2]Department of Physics, Tokyo Metropolitan University, Tokyo 192-0397, Japan

Corresponding Author
*E-mail: zaumseil@uni-heidelberg.de

ABSTRACT

The covalent functionalization of single-walled carbon nanotubes (SWCNTs) with luminescent quantum defects enables their application as near-infrared single-photon sources, as optical sensors, and for *in-vivo* tissue imaging. Tuning the emission wavelength and defect density are crucial for these applications. While the former can be controlled by different synthetic protocols and is easily measured, defect densities are still determined as relative rather than absolute values, limiting the comparability between different nanotube batches and chiralities. Here, we present an absolute and unified quantification metric for the defect density in SWCNT samples based on Raman spectroscopy. It is applicable to a range of small-diameter nanotubes and for arbitrary laser wavelengths. We observe a clear inverse correlation of the D/$G^+$ ratio increase with nanotube diameter, indicating that curvature effects contribute significantly to the defect-activation of Raman modes. Correlation of intermediate frequency modes with defect densities further corroborates their activation by defects and provides additional quantitative metrics for the characterization of functionalized SWCNTs.

Semiconducting single-walled carbon nanotubes (SWCNTs) with luminescent defects introduced by sp³ functionalization have become an attractive material platform for bioimaging,[1,2] optical sensing,[3-5] electroluminescent devices,[6,7] and single-photon sources[8,9] in the near-infrared (nIR), especially beyond 1000 nm. These quantum defects are typically created by covalent functionalization of the sp²-hybridized carbon lattice of SWCNTs with aryl or alkyl groups resulting in few sp³ carbon atoms along the nanotube and corresponding changes in the local electronic structure.[10] The defects form energetically lower lying emissive states that act as zero-dimensional traps for mobile excitons, leading to further red-shifted narrowband photoluminescence (PL) compared to pristine nanotubes and PL lifetimes of hundreds of picoseconds.[11,12] Due to the localization of previously mobile excitons, the overall PL quantum yield (PLQY) can increase substantially[13] and room-temperature single-photon emission can be observed.[8,9]

The emission wavelength of luminescent sp³ defects is largely determined by the nanotube (n,m) chirality and the binding configuration of the defect on the nanotube lattice.[11,14] The latter can be selected by the choice of the functionalization method.[9,15,16] The degree of functionalization and hence the number of defects per nanotube length can be tuned by the reaction time or concentration of reactants and must be adjusted depending on the intended application. With ongoing efforts to utilize sp³-functionalized SWCNTs for a variety of applications ranging from cancer detection to quantum information processing,[4,17] the ability to reliably and easily determine the precise density of sp³ defects in a nanotube sample has become crucial to characterize and control their impact on the optical and electronic properties of SWCNTs.

For graphene, a method to determine the number density of point-like structural defects is well-established and has even been expanded to line defects.[18,19] It is based on Raman spectroscopy, which is highly sensitive to structural deformations of the carbon lattice. To obtain a quantitative model, Raman spectra were cross-correlated with the number of defects introduced by ion bombardment as determined by scanning tunneling microscopy.[20] A higher defect density was accompanied by a larger D-mode intensity (originating from second-order Raman scattering processes involving one phonon and elastic scattering at a defect site) relative to the G-mode (proportional to the number of $sp^2$-hybridized carbon atoms). The resulting empirical expression for the average defect distance $L_d$ in graphene also includes the dependence of the observed D/G intensity ratio $\left(\frac{I_D}{I_G}\right)$ on the excitation laser wavelength $\lambda_L$;[18]

$$L_d^2(nm^{-2}) = (1.8 \pm 0.5) \times 10^{-9} \lambda_L^4 \cdot \left(\frac{I_D}{I_G}\right)^{-1} \tag{1}$$

This simple metric enables the absolute quantification of defects in graphene and has been used frequently to determine the quality after exfoliation[21] but also the degree of covalent functionalization.[22]

In contrast to that, the development of a universal metric to quantify the number of defects in SWCNTs has proven difficult. Similar to graphene, second-order Raman scattering processes at defect sites give rise to a D-mode around 1320 cm$^{-1}$. Its intensity relative to the $G^+$ optical phonon mode of the $sp^2$-hybridized carbon lattice is commonly used as a relative indicator for the presence of defects.[23,24] The wavenumber of the D-mode in SWCNTs also shows a dependence on excitation laser energy and nanotube diameter.[25-27] Although it was speculated that the quantification metric developed for graphene might also be applicable to other $sp^2$-hybridized carbon materials,[28] only few attempts were made to systematically correlate D/$G^+$ ratios from Raman spectra of SWCNTs with defect densities obtained by independent measurements.[29-31] Compared to the atomically defined monolayers of graphene produced by chemical vapor deposition or mechanical exfoliation, the controlled introduction and direct real-space evaluation of structural defects in SWCNTs is difficult due to their cylindrical geometry and bundling effects in bulk samples. Furthermore, the purification of SWCNT raw material is usually carried out in the liquid phase and yields a distribution of different chiralities, diameters, and lengths depending on the specific protocol,[32,33] which have additional effects on the measured Raman modes.

Recently, we have shown that a linear correlation between the integrated Raman $D/G^+$ mode ratio at a fixed excitation wavelength and defect densities obtained from absolute PLQY measurements of pristine and covalently functionalized SWCNTs can be used to directly quantify the number of introduced $sp^3$ defects in (6,5) SWCNTs.[34] While this quantification method was shown to be reliable for (6,5) nanotubes in different types of dispersion (organic solvent or aqueous), for different lengths of nanotubes and even for different types of $sp^3$ defects, it remained unclear if it could also be applied to other SWCNT chiralities and diameters as well as for different Raman excitation wavelengths similar to graphene.

Here, we introduce a unified empirical expression for the quantification of $sp^3$ defects in small-diameter (< 1 nm) semiconducting SWCNTs using the $D/G^+$ ratios obtained from Raman spectra collected at arbitrary excitation wavelengths. We find a dependence of the linear calibration factor on the nanotube diameter, which indicates a strong impact of curvature and strain on defect-activated Raman modes. This generalized approach provides a noninvasive and fast way to quantify the absolute number of $sp^3$ defects in various SWCNTs using only Raman spectroscopy. In addition to the familiar D-mode, we also consider other defect-activated Raman modes in the intermediate frequency range of SWCNTs as quantitative metrics for the $sp^3$ functionalization of nanotubes.

RESULTS AND DISCUSSION

**SWCNT functionalization and quantification of sp$^3$ defects**

Monochiral dispersions of small-diameter, semiconducting SWCNTs were either prepared by selective polymer-wrapping in toluene or by gel-chromatography of SWCNTs stabilized by surfactants in water. To prepare dispersions of (6,5) and (7,5) SWCNTs, CoMoCAT raw material was dispersed by shear force mixing in toluene solutions of different polyfluorene-based polymers.[35] Aqueous dispersions of (7,3), (8,3), and (9,2) SWCNTs were obtained from HiPco raw material employing a previously reported procedure based on a high-performance liquid chromatography (HPLC) system (for details, see **Experimental Methods**).[36] Prior to sp$^3$ functionalization, the purity of the pristine SWCNT dispersions was assessed by UV-Vis-nIR absorption spectroscopy, PL excitation-emission mapping and Raman spectroscopy. Absorption spectra of pristine SWCNT dispersions (see **Supporting Information, Figure S1**) show the characteristic $E_{11}$ and $E_{22}$ optical transitions along with the respective phonon side bands but essentially no indication of any other SWCNT chiralities except for residual (7,6) SWCNTs in the (7,5) nanotube dispersion. The PL excitation-emission maps in **Figure S2** (**Supporting Information**) show the characteristic $E_{11}$ emission at $E_{22}$ excitation for each purified nanotube chirality and no additional semiconducting SWCNT species. The (n,m) chiral index assignment and absence of metallic carbon nanotubes was further confirmed using averaged Raman spectra in the region of the radial breathing modes (RBM) at excitation wavelengths of 532, 633, and 785 nm (see **Supporting Information, Figure S3**).[37,38] Lorentzian fits of Raman peaks in the region from 1500 to 1700 cm$^{-1}$ (see **Supporting Information, Figure S4**) reveal the characteristic diameter-dependence of the G$^+$ and G$^-$ mode frequencies in agreement with the model by Telg *et al*.[39]

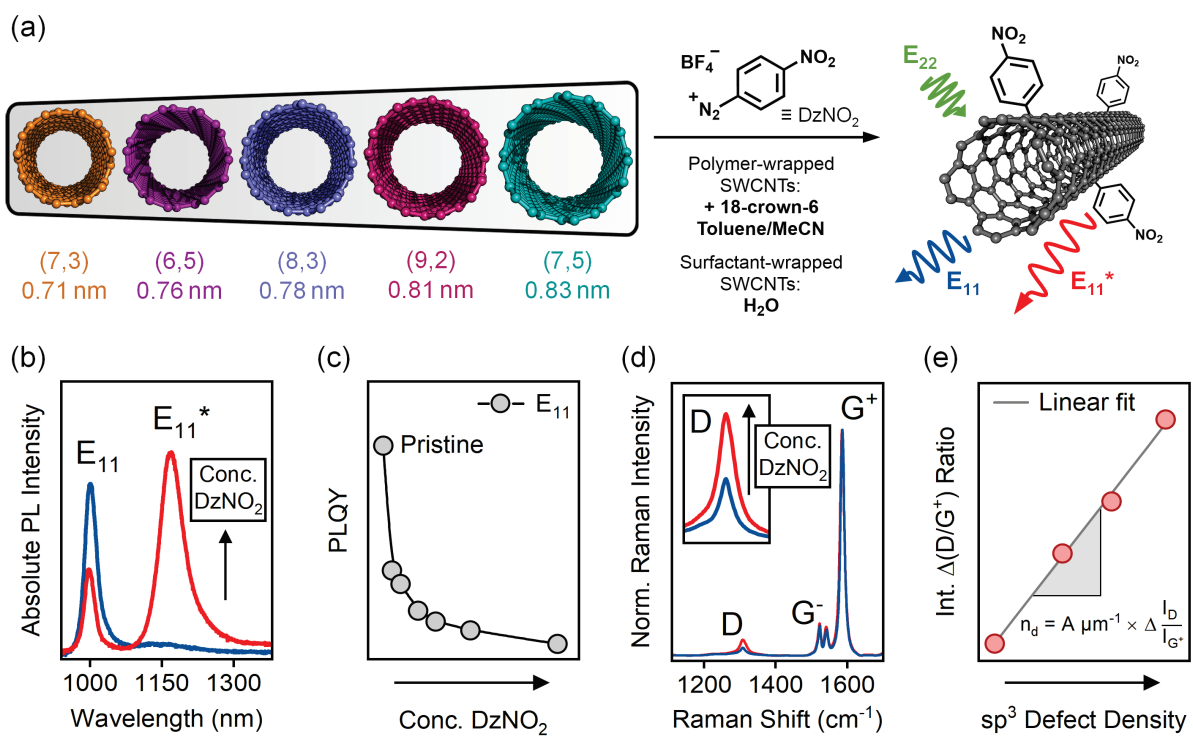

**Figure 1. (a)** SWCNT chiralities with different diameters that are investigated here and reaction scheme for sp$^3$ functionalization using 4-nitrobenzenediazonium tetrafluoroborate (DzNO$_2$) in different solvent systems. Luminescent sp$^3$ defects act as trapping sites for mobile excitons created by $E_{22}$ excitation, giving rise to redshifted $E_{11}$* emission in addition to $E_{11}$ emission. **(b)** Exemplary PL spectra of sp$^3$-functionalized (6,5) SWCNTs (red) showing $E_{11}$* emission compared to pristine nanotubes (blue). **(c)** PLQY of only the $E_{11}$ emission for pristine and sp$^3$-functionalized (6,5) SWCNTs. The line is a guide to the eye. **(d)** Normalized average Raman spectra of pristine and sp$^3$-functionalized (6,5) SWCNTs and assignment of characteristic D, G$^-$, and G$^+$ modes, with inset showing the D-mode. **(e)** Correlation of integrated differential Raman Δ(D/G$^+$) ratios with sp$^3$ defect densities calculated from $E_{11}$ PLQY and linear fit to the data.

The different pristine SWCNTs were covalently functionalized with luminescent sp$^3$ defects using 4-nitrobenzenediazonium tetrafluoroborate (DzNO$_2$) as shown in **Figure 1a**, and the degree of sp$^3$ functionalization was tuned by adjusting the diazonium salt concentration. The functionalization of SWCNT dispersions in water was achieved by simply adding an aqueous solution of DzNO$_2$,[13] whereas the functionalization of polymer-wrapped nanotubes in toluene was aided by 18-crown-6 as a phase-transfer reagent.[40] The appearance of an additional red-shifted emission (E$_{11}$*) in the PL spectra of functionalized SWCNTs confirmed the introduction of luminescent sp$^3$ defects, as shown exemplarily for (6,5) SWCNTs in **Figure 1b**. While the total PLQY of the SWCNT dispersions at low-to-medium functionalization levels increased due to efficient trapping of highly mobile excitons at sp$^3$ defects and their radiative decay, the PLQY related only to the E$_{11}$ emission always decreased (see **Figure 1c**). From this decrease of E$_{11}$ PLQY compared to pristine nanotubes, the number density of luminescent sp$^3$ defects was calculated using the diffusion-limited contact quenching (DLCQ) model for excitons in SWCNTs.[41] Mobile excitons either decay radiatively through E$_{11}$ emission or undergo nonradiative quenching at nanotube ends or lattice defects.[42,43] In the presence of luminescent sp$^3$ defects, the E$_{11}$ PLQY is also reduced as excitons are trapped and recombine at the sp$^3$ defects, leading to E$_{11}$* PL emission. An analytical expression for the density of introduced luminescent sp$^3$ defects ($n_d$ in µm$^{-1}$) independent of the number of initial quenching sites can be derived based on the ratio of the E$_{11}$ PLQY of pristine and sp$^3$-functionalized SWCNTs ($\eta$ and $\eta$*, respectively, for details of the calculation and PLQY measurements see **Experimental Methods**).

The calculated sp$^3$ defect densities also depend on the chosen values for the exciton diffusion constant $D$ and the E$_{11}$ radiative lifetime $\tau_{rad}$. Reported values for $D$ are typically within a range of 2-10 cm$^2$ s$^{-1}$,[43-45] whereas $\tau_{rad}$ was determined to be on the order of a few nanoseconds.[46,47] While there are only few studies on the effect of nanotube diameter and chirality on $D$ and $\tau_{rad}$, the available data consistently suggest no significant variations or clear trends in the investigated range of nanotube diameters.[47-49] Thus, $D$ and $\tau_{rad}$ as determined for (6,5) SWCNTs in previous experimental studies[42,46] were employed for all calculations of sp$^3$ defect densities irrespective of nanotube chirality. Independent counting of defect emission peaks from single nanotubes at cryogenic temperatures for different batches of functionalized (6,5) nanotubes previously indicated that the defect densities obtained with the described DLCQ method were indeed very close to the average numbers within the nanotube ensemble.[34]

**Raman spectroscopy of functionalized SWCNTs**

The defect density in SWCNTs is also reflected in their Raman spectra, where an increase of the $D/G^+$ mode intensity ratio indicates more defects, as shown in **Figure 1d**. For purified samples and within the limit of moderate functionalization, *i.e.*, without clustering of defects, the intensity of the Raman D-mode is proportional to the number of $sp^3$-hybridized carbon atoms.[50] As recently demonstrated for (6,5) SWCNTs, the difference of the integrated $D/G^+$ ratio between pristine and functionalized samples ($\Delta(D/G^+)$) can be used as an absolute metric for the number of introduced defects based on its linear correlation with the $sp^3$ defect densities obtained from $E_{11}$ PLQY measurements (see **Figure 1e**).[34] It is independent of the initial number of defects in the raw material, the length of the nanotubes or the dispersion medium.

To expand this approach to other SWCNT chiralities and to investigate its dependence on the Raman excitation wavelength and nanotube diameter, pristine and functionalized (7,3), (6,5), (8,3), (9,2), and (7,5) SWCNTs were characterized by PL and Raman spectroscopy. For all nanotube species, the $sp^3$ functionalization resulted in emission from $E_{11}$* defects as observed in the PL spectra of the SWCNT dispersions (see **Supporting Information, Figures S5a - S9a**). Note that the relative PL intensities of defect and $E_{11}$ emission are heavily influenced by the excitation power and the dielectric environment of the nanotubes (*e.g.*, surfactant and solvent) and do not reflect the actual density of defects.[16] Spectrally resolved PLQY measurements reveal a sharp decrease of the absolute $E_{11}$ PLQY with increasing $sp^3$ functionalization for all nanotube species and enable the direct calculation of defect densities based on the DLCQ model as outlined above (see **Supporting Information, Figures S5b - S9b**).

For the correlation of the calculated sp$^3$ defect densities $n_d$ and the D/G$^+$ intensity ratios, Raman maps of dropcast films of pristine and sp$^3$-functionalized SWCNTs were collected at excitation wavelengths of 532, 633, and 785 nm, and over 3000 individual spectra were averaged for each sample. The integrated D and G$^+$ mode intensities were obtained from Lorentzian fits to the averaged and baseline-corrected Raman spectra. The results for (9,2) and (7,3) SWCNTs are shown exemplarily in **Figure 2** (for other SWCNT chiralities see **Supporting Information, Figures S10-S13**). The dispersive nature of the Raman D-mode[26] becomes evident for longer excitation wavelengths as the peak maximum shifts to lower wavenumbers. We observe no significant change in peak position or width of the G$^+$ mode for pristine and sp$^3$-functionalized SWCNTs, indicating that the perturbation of the sp$^2$-hybridized carbon lattice and the SWCNT band structure for moderate sp$^3$ functionalization is still negligible.

In addition to the expected increase of the Raman D/G$^+$ intensity ratio with the concentration of DzNO$_2$ during functionalization, the D/G$^+$ intensity ratio also shows a strong dependence on the excitation wavelength. For Raman spectra recorded with longer excitation wavelengths $\lambda_L$ and thus lower excitation energies $E_L$, the D-mode becomes much more pronounced compared to the G$^+$ mode. This effect was initally observed and quantified in Raman studies of nanographites and graphene,[18,51] and can be explained by the proportionality of the Raman cross section of the G-mode to the excitation energy $E_L^4$. In contrast to that, the cross section of the D-mode does not show any dependence on $E_L$. Systematic studies of this effect in SWCNTs were previously restricted to ion-bombarded double-walled nanotubes due to the limited availability of monochiral dispersions at the time. However, they indicated the same dependence.[28,52]

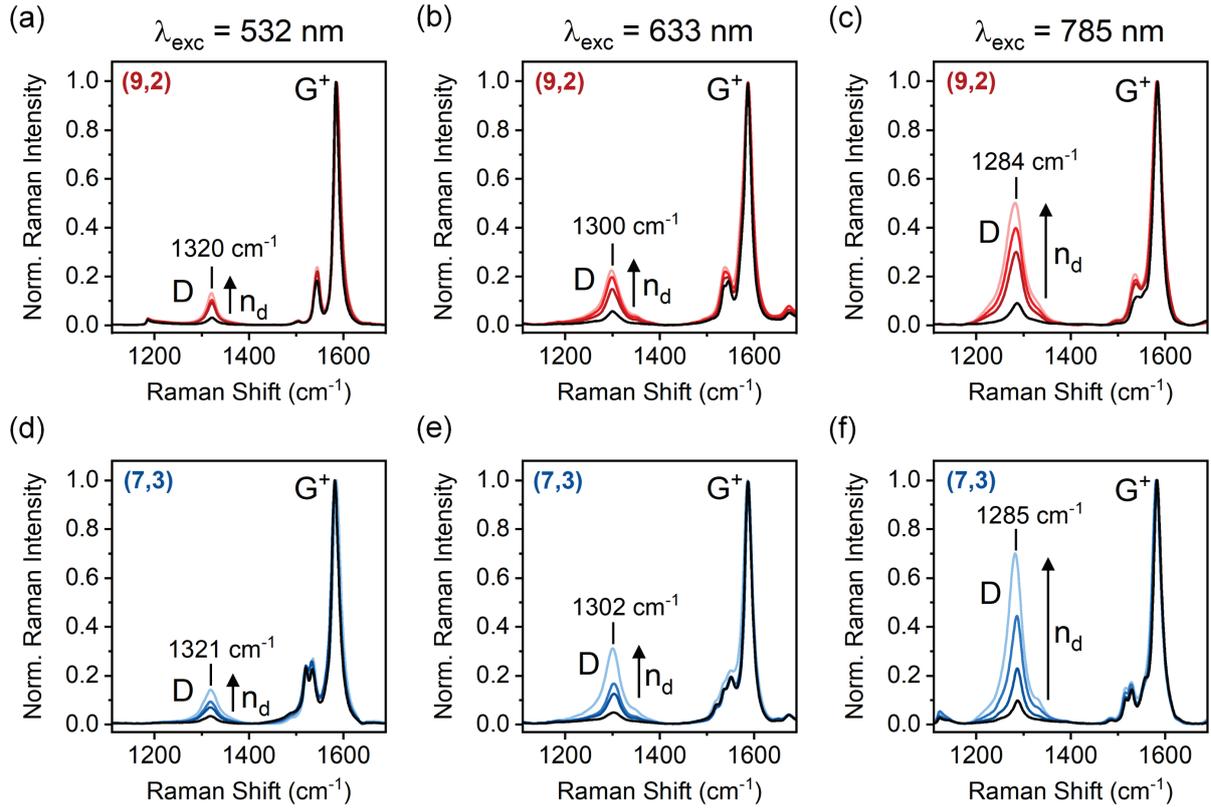

**Figure 2.** Averaged (>3000 spectra for each sample) and normalized Raman spectra of pristine and $sp^3$-functionalized (9,2) **(a-c)** and (7,3) SWCNTs **(d-f)**, acquired at excitation laser wavelengths of 532 nm ($E_L$ = 2.33 eV) **(a, d)**, 633 nm ($E_L$ = 1.96 eV) **(b, e)**, and 785 nm ($E_L$ = 1.58 eV) **(c, f)**, demonstrating the red-shift of the Raman D-mode and higher D/G$^+$ intensity ratios with lower excitation energies (longer excitation wavelengths).

**Laser wavelength-independent defect quantification**

To evaluate the transferability of the model derived for point-like defects in nanographites and graphene to SWCNTs, we correlate the Raman D/G$^+$ ratio with the calculated $sp^3$ defect density $n_d$ from PLQY measurements for all investigated SWCNT chiralities and excitation wavelengths. For this quantitative analysis, the difference between the integrated D/G$^+$ ratios (*i.e.,* the area ratios) instead of the intensity ratios is used, as it provides a more robust measure when only small perturbations of the pristine SWCNT lattice are present.[18] Furthermore, the integrated Raman D/G⁺ ratio for the pristine SWCNTs is subtracted from all sp³-functionalized samples, resulting in a differential value, the Δ(D/G⁺) ratio. This differential value enables the application of this analysis to SWCNT samples with different initial defect densities and hence D-mode intensities before functionalization.

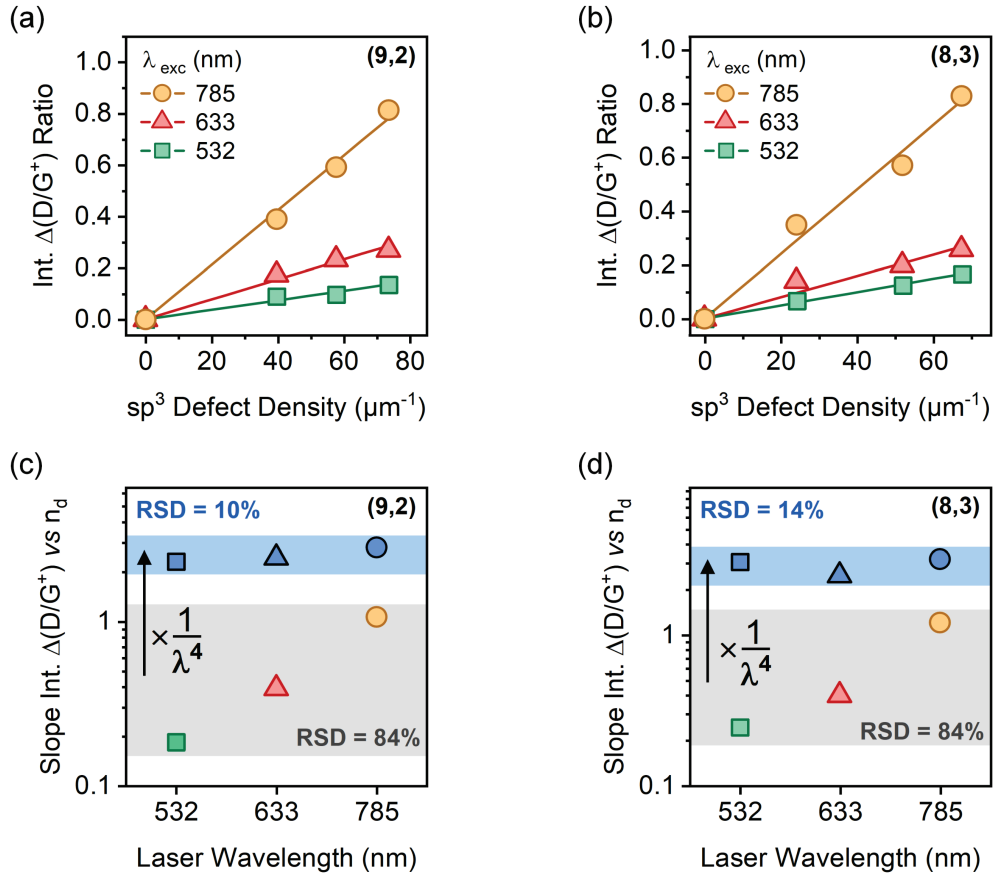

**Figure 3.** Integrated Raman Δ(D/G⁺) ratio *vs* calculated sp³ defect density $n_d$ for pristine and sp³-functionalized (9,2) **(a)** and (8,3) SWCNTs **(b)** at laser wavelengths of 532 nm ($E_L$ = 2.33 eV), 633 nm ($E_L$ = 1.96 eV), and 785 nm ($E_L$ = 1.58 eV), including linear fits to the data ($R^2$ (532 nm) = 0.99, $R^2$ (633 nm) = 0.99, $R^2$ (785 nm) = 0.99). (b) Slopes of linear fits in (a) and (b) for different excitation wavelengths (*i.e.*, excitation energy $E_L$) and their normalization with $E_L^4$ or $1/\lambda^4$ for **(c)** (9,2) and **(d)** (8,3) SWCNTs including the relative standard deviation (RSD in %) before and after normalization.

**Figures 3a** and **3b** display the obtained plots of the integrated Δ(D/G$^+$) ratios *vs* calculated defect densities $n_d$ for (9,2) and (8,3) SWCNTs, respectively (for data on other chiralities, see **Supporting Information, Figures S13a - S15a**). In all cases, a linear correlation between the Δ(D/G$^+$) ratio and the sp$^3$ defect density $n_d$ can be found, but with very different slopes depending on the excitation wavelengths (532, 633, and 785 nm). Longer excitation wavelengths lead to greater slopes. The values for the slopes of Δ(D/G$^+$) *vs* $n_d$ and the respective standard deviations obtained from linear fitting are reported in **Table 1** for all investigated SWCNT chiralities.

In direct analogy to graphene,[18] a normalization of the slopes (here Δ(D/G$^+$) *vs* $n_d$) by the laser excitation wavelength $\lambda_L^4$ (or interchangeably to the excitation energy $E_L^{-4}$) leads to unified values with a substantially reduced relative standard deviation (RSD), as shown in **Figures 3c** and **3d** (for other chiralities, see **Supporting Information, Figures S13b-S15b**). The wavelength-corrected slopes for Δ(D/G$^+$) *vs* $n_d$ can be averaged to yield a universal empirical expression for the sp$^3$ defect density (with $\lambda_L$ in nanometers):

$$n_d(\mu m^{-1}) = \frac{A(d_t) \cdot 10^{13}}{\lambda_L^4} \cdot \Delta\left(\frac{I_D}{I_{G^+}}\right) \qquad (2)$$

The resulting values for the calibration factor *A* are reported in **Table 1** for each SWCNT chirality. Equation (2) is applicable to SWCNT chiralities within the diameter range investigated here (0.7 – 0.85 nm) and for sp$^3$ defect densities in the range of ~5-80 μm$^{-1}$ with an estimated accuracy of ±15%. Note that maximum PLQYs and highest defect-to-E$_{11}$ PL intensities of functionalized nanotubes are typically obtained at ~6-10 luminescent sp$^3$ defects per μm of SWCNT.[40] Thus, the presented quantification method can be used to guide the selection of suitable samples with optimum E$_{11}$* defect emission in optoelectronic devices or nanotube-based sensors based on simple Raman spectroscopy. It could further ensure comparability between different batches and sources of carbon nanotubes. Here we used both CoMoCat and HiPco nanotubes as well as different types of dispersion (polymer-wrapped in organic solvents and surfactant-stabilized in aqueous environments, see **Methods**) and still obtained a continuous linear correlation.

**Table 1.** Slopes (standard deviations in brackets) of the integrated $\Delta(D/G^+)$ ratios *vs* the calculated defect density $n_d$ for all investigated (n,m) chiralities of SWCNTs at three different Raman laser excitation wavelengths, nanotube diameter $d_t$, and calibration factor $A(d_t)$ (standard deviation) for each SWCNT chirality as used in eq. (2).

| Laser wavelength (nm) | (7,3) | (6,5) | (8,3) | (9,2) | (7,5) |
|---|---|---|---|---|---|
| 532 | 0.0038 (0.0004) | 0.0028 (0.0002) | 0.0025 (0.0001) | 0.0019 (0.0002) | 0.0014 (0.0002) |
| 633 | 0.0087 (0.0006) | 0.0073 (0.0003) | 0.0040 (0.0003) | 0.0039 (0.0003) | 0.0025 (0.0001) |
| 785 | 0.019 (0.002) | 0.0186 (0.0007) | 0.0120 (0.0006) | 0.0106 (0.0003) | 0.0088 (0.0003) |
| $d_t$ *(nm)* | 0.71 | 0.76 | 0.78 | 0.81 | 0.83 |
| $A(d_t)$ | 2.0 (0.2) | 2.4 (0.4) | 3.5 (0.5) | 4.0 (0.4) | 5.4 (1.1) |

**Dependence of D/G$^+$ ratios on nanotube diameter**

A closer look at the calibration factor *A* for the different SWCNT chiralities reveals a clear dependence on nanotube diameter ($d_t$) as shown in **Figure 4a**. For example, at identical calculated defect densities, the D/G$^+$ ratio for (7,3) SWCNTs ($d_t$ = 0.71 nm) is larger by a factor of ~2.5 compared to (7,5) SWCNTs ($d_t$ = 0.83 nm). At first glance, this deviation could be explained by a larger relative impact of individual sp$^3$ defects on the D-mode of SWCNTs with smaller diameters. To rationalize this argument, we adopt the model of D-mode activation in the lattice area directly around point-like defects, as developed for graphene.[20] Within this picture, a single sp³ defect may result in a larger relative increase of the D-mode intensity when the number of sp²-hybridized carbon atoms contributing to the $G^+$ mode is lower. This would be the case for SWCNTs with smaller diameters and hence circumference but identical length. This effect can be quantified by unfolding the carbon nanotube lattice to obtain a two-dimensional graphene sheet with a defined area, as schematically depicted in **Figure 4b**. The number of carbon atoms per unit length of each nanotube chirality can thus be used to normalize the calibration factors *A* (see **Supporting Information, Table S1** and **Figure S16**). While this normalization reduces the deviation between different SWCNT chiralities, nanotubes with smaller diameters still show significantly higher $D/G^+$ ratios for the same sp³ defect density (*i.e.*, smaller *A*) compared to SWCNTs with larger diameters. Evidently, for the investigated nanotube diameters, the number of carbon atoms per μm for different nanotubes cannot fully explain the diameter dependence of *A*.

In addition to the pronounced diameter dependence of the $D/G^+$ ratio, other possible correlations were tested. As each sp³ defect involves two sp³ carbons in different ortho or para configurations,[14] their alignment with respect to the nanotube axis might also be reflected in the $D/G^+$ ratio, although at least for (6,5) nanotubes no dependence on the binding configuration was observed.[34] Among the investigated SWCNT chiralities, we find no distinct correlation with chiral angle or mod family (mod(2n+m , 3) = 1 or 2) as shown in **Figure S17** (**Supporting Information**). However, at similar chiral angles, a lower slope of $\Delta D/G^+$ versus $n_d$ is observed for mod family 2 compared to mod family 1 nanotubes. A comprehensive study of a broader range of SWCNTs with similar diameters but different chiral angles or belonging to different mod families could resolve possible dependencies but is beyond the scope of this work.

Based on the available data, curvature effects appear to play a major role in highly-strained SWCNTs with diameters below 1 nm. For the wavenumber of the D-mode in SWCNTs, an inverse dependence on nanotube diameter was determined experimentally and supported by numerical simulations.[26,27] In particular, strong deviations from the tight-binding and zone-folding model emerge in the dispersion of the transverse optical (TO) phonons at the K point of the Brillouin zone in small-diameter SWCNTs.[25,53] A point-defect in a highly strained lattice might be expected to have a larger impact than in a less strained or flat $sp^2$ carbon lattice despite the fact that there is already a higher $sp^3$ character in pristine nanotubes with small diameters than in large-diameter nanotubes.[54]

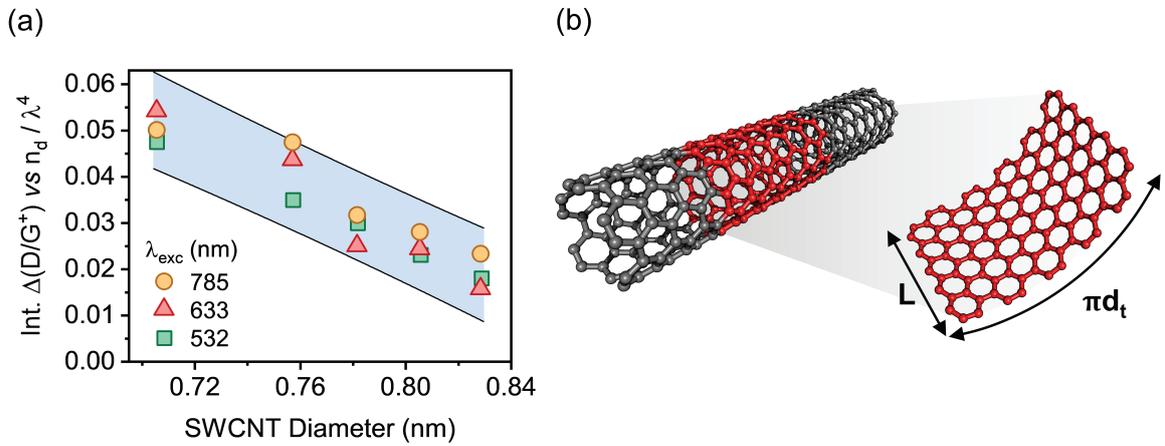

**Figure 4.** (a) Integrated Raman $\Delta(D/G^+)$ vs $n_d$ ratios for different diameters of SWCNTs, normalized to the excitation laser wavelength $\lambda_L^4$. (b) Schematic illustration of the conceptual unfolding of the nanotube lattice for the normalization of the $D/G^+$ ratio to the SWCNT area ($L \cdot \pi d_t$), where $L$ denotes the length of the nanotube segment and $d_t$ the tube diameter.

If the diameter dependence of the relative increase of the $D/G^+$ ratio versus defect density is related to curvature effects in SWCNTs, an expansion of the one-dimensional nanotube to a two-dimensional graphene sheet (equivalent to $d_t \to \infty$) may result in correlations similar to graphene. By representing the average defect density in nanometer via $L_d(nm) = 1000 \, nm/n_d \, (\mu m^{-1})$, a direct comparison between the empirical expressions obtained for graphene (eq. (1)) and SWCNTs (eq. (2)) can be drawn. We find that the $D/G^+$ ratios for a given defect density below ~20 $\mu m^{-1}$ as calculated with eq. (1) are on the same order of magnitude as those determined experimentally for the different nanotube chiralities (see **Supporting Information, Figure S18 and Table S2**). The investigated nanotubes with larger diameters (*i.e.*, (9,2) and (7,5)) are closer to the values calculated for graphene. But smaller diameter nanotubes show higher $D/G^+$ ratios for the same low defect density. However, for very high defect densities (>30 µm$^{-1}$) the $D/G^+$ ratios for graphene deviate substantially and become consistently higher than the experimental values for nanotubes. It is not clear, how these numbers can be reconciled. Clearly, rolling-up a planar graphene sheets into small-diameter, highly-strained SWCNTs, exerts a strong influence on the relative ratio of the Raman D and $G^+$ modes for a given number of defects and the empirical expression for graphene cannot be applied to quantify defects in carbon nanotubes.

Although no conclusive quantitative relation between the nanotube diameter and changes in the $D/G^+$ ratios with defect density can be drawn from the limited range of five small-diameter SWCNTs, the available data strongly indicates that caution has to be taken when comparing defect densities of different carbon nanotube chiralities. Samples with different diameter distributions will give rather different $D/G^+$ ratios for the same or similar defect densities (initial or introduced) in addition to a strong dependence on the Raman laser wavelength. Future defect density analyses would profit greatly from more detailed theoretical studies on the impact of nanotube geometry as well as defect configuration on Raman D and $G^+$ modes in functionalized or defective SWCNTs.

**Activation of intermediate frequency modes by sp$^3$ defects**

So far, we have only considered the impact of sp$^3$ defects on the D-mode of SWCNTs, which is arguably the most prominent indication of defects. However, additional defect-induced Raman features can be found in the spectral range between the RBM and D-mode, which are consequently named intermediate frequency modes (IFMs).[55] These IFMs typically exhibit much lower intensities than the D-mode and are most pronounced in Raman spectra of individual aligned SWCNTs.[56] Among the different IFMs, the signature of the ZA-derived phonon branch of graphene occurs in the range between 450 and 500 cm$^{-1}$ and was found to increase with higher levels of defects.[56-60] As expected for a defect-related double-resonant process, the ZA-derived mode exhibits dispersive characteristics.[61] Its intensity also increases with longer excitation wavelengths. An additional Raman signal at ~600 cm$^{-1}$ was also proposed to originate from a defect-induced double-resonant process involving the ZO-derived phonon branch.[56] Recent theoretical studies based on excitation close to E$_{11}$ have further indicated the possible presence of defect-related modes in the IFM region.[62]

To clarify the nature of IFM features and their correlation with sp$^3$ defect densities of small-diameter nanotubes, we collected averaged Raman spectra in the IFM region (excitation wavelength 785 nm) for all pristine and sp$^3$-functionalized SWCNT samples. The Raman spectra for (6,5) SWCNTs normalized to the nearby RBM peak are shown in **Figure 5a** (for additional chiralities see **Supporting Information, Figures S19a-S22a**). ZA- and ZO-derived phonon modes at 480 cm$^{-1}$ and 570 cm$^{-1}$, respectively, are evident and increase in intensity relative to the RBM signal with sp$^3$ defect density. Hence, they could be used in the same way as the D/G$^+$ ratio for defect quantification as discussed above. We determined the integrated differential ratio Δ(IFM/RBM) from pristine and functionalized SWCNTs and found a linear correlation with the sp$^3$ defect density $n_d$ for both IFMs (see **Figure 5b** and **Supporting Information, Figures S19b-S22b**). Due to the highly selective and systematic introduction of sp$^3$ defects, we can conclude that both the ZA- and the ZO-derived phonon modes are indeed the result of defect-activated, double-resonant scattering processes as previously suggested.[56]

The slopes and respective standard deviations for linear fits of Δ(IFM/RBM) *vs* n$_d$ are reported in **Table S3** (**Supporting Information**). They show strong variations with SWCNT species, which is most likely the result of the complex chirality and mod family dependence of the RBM to G$^+$ mode intensities, as reported previously for a variety of SWCNT chiralities.[63] In addition, the analysis of the IFM region is often complicated by traces of other SWCNT chiralities that also contribute to Raman features in the investigated spectral region resulting in larger errors of the linear fits for (7,3) and (9,2) SWCNTs. Nevertheless, our findings indicate that for highly pure SWCNT samples, defect-related IFM features can also be used for the absolute quantification of sp$^3$ defects and hence offer an additional metric for covalent nanotube functionalization.

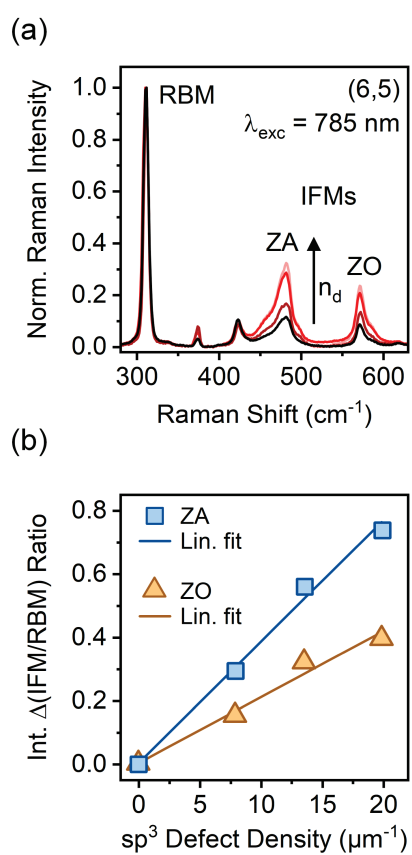

**Figure 5. (a)** Averaged and normalized (to RBM) Raman spectra ($\lambda_{exc}$ = 785 nm) of pristine and sp$^3$-functionalized (6,5) SWCNTs in the intermediate frequency mode (IFM) region. **(b)** Integrated Δ(IFM/RBM) intensity ratio for the IFMs depicted in (a) and linear fits to the data ($R^2$ = 0.99 for ZA mode, and $R^2$ = 0.99 for ZO mode).

CONCLUSION

We have demonstrated a combined analysis of Raman spectra and PLQY measurements of functionalized SWCNTs for the absolute quantification of sp$^3$ quantum defects across different nanotube chiralities and diameters. Covalently functionalized (7,3), (6,5), (8,3), (9,2), and (7,5) SWCNTs with low- to intermediate levels of sp$^3$ defects showed a linear correlation of the defect densities obtained from $E_{11}$ PLQY measurements with the differential increase of the integrated Raman D/G$^+$ intensity ratios. A normalization of the resulting slopes to the Raman excitation wavelength $\lambda_L^4$ leads to a simple, excitation energy-independent linear expression for the defect density in SWCNTs and thus enables a straightforward quantification of the number of introduced defects in nanotubes from Raman spectra alone. An inverse correlation between the integrated D/G$^+$ ratio increase and the nanotube diameter $d_t$ indicates that curvature effects and strain exert a strong influence on defect-activated Raman modes in SWCNTs. This fairly strong diameter dependence of the D/G$^+$ ratio for different nanotube diameters also demonstrates that simple comparisons between different nanotube batches should be applied with caution. In addition to the characteristic increase of the D-mode intensity, sp$^3$-functionalization is also reflected in the intermediate frequency Raman modes of SWCNTs. Both the ZA- and ZO-derived phonon modes can be clearly assigned as defect-activated as their relative intensities scale linearly with defect density. They could be used as an additional and complementary metric for the degree of nanotube functionalization.

These unified defect characterization and quantification methods based on simple Raman spectroscopy should enable a more systematic evaluation of the efficiency of SWCNT functionalization methods as well as quality control for a range of nanotubes with sp$^3$ defects that find application in biosensing and optoelectronics. They should also incite further systematic studies of a broader range of nanotube chiralities with different types of lattice defects, including for example oxygen defects. Finally, the development of further theoretical descriptions is required to arrive at a comprehensive picture and quantitative model for defect-related effects in SWCNT Raman spectra.

EXPERIMENTAL METHODS

**Preparation of Monochiral SWCNT Dispersions.** (6,5) and (7,5) SWCNTs: Dispersions of (6,5) SWCNTs were obtained by shear force mixing (SFM, Silverson L2/Air, 10,230 rpm, 20 °C, 72 h) of CoMoCAT raw material (Sigma-Aldrich, Charge No. MKCJ7287) in a solution of poly[(9,9-dioctylfluorenyl-2,7-diyl)-*alt*-(6,6′-(2,2′-bipyridine))] (PFO-BPy, American Dye Source, $M_w$ = 40 kg·mol$^{-1}$, 0.5 g·L$^{-1}$) in toluene as reported previously.[35] (7,5) SWCNTs were likewise dispersed by shear force mixing in a solution of poly[(9,9-dioctylfluorenyl-2,7-diyl)] (PFO, Sigma-Aldrich, $M_w$ >20 kg·mol$^{-1}$, 0.9 g·L$^{-1}$) in toluene. Unexfoliated material was removed by centrifugation (Beckman Coulter Avanti J26XP centrifuge) at 60000 *g* for 1 hour and filtration of the supernatant through a poly(tetrafluoroethylene) (PTFE) syringe filter (pore size 5 μm). Vacuum filtration through PTFE membrane filters (Merck Millipore JVWP, pore size 0.1 μm) and washing of the resulting filter cakes with hot toluene (80 °C, 10 min) was performed to remove excess polymer, followed by redispersion of the polymer-wrapped nanotubes in fresh toluene *via* bath sonication.

(7,3), (8,3), and (9,2) SWCNTs: Monochiral (7,3), (8,3), and (9,2) SWCNTs were obtained by gel chromatography employing a mixed surfactant system following a previously reported protocol.[36,64] Initially, HiPco SWCNT raw material (NanoIntegris, Inc., diameter distribution 1.0 ± 0.3 nm) was dispersed in an aqueous solution of sodium cholate (SC, 0.8 w/vol.-%, >99%, Sigma-Aldrich) by ultrasonication (Branson Sonifier 250D) with 30% output power for 1-2 hours. After ultracentrifugation at 210000 *g* for 2 hours, the collected supernatant was mixed with aqueous solutions of sodium dodecyl sulfate (SDS, >99%, Sigma-Aldrich) and SC to adjust the surfactant concentration depending on the (n,m) chirality. The separation was carried out with a high-performance liquid chromatography system (Cytiva AKTAexplorer 10S) and columns (Cytiva HiScale 26/20) filled with gel beads (Cytiva Sephacryl S-200 HR) at a temperature of 17 °C. The surfactant concentration of the SWCNT dispersion was adjusted to 0.6% SDS + 0.4% SC. After equilibration of the column with 0.6% SDS + 0.4% SC and loading with the SWCNT dispersion, adsorbed SWCNTs were eluted using a mixed surfactant solution of 0.6% SDS + 0.4% SC + x% sodium lithocholate (LC, >98%, Tokyo Chemical Industry Co., Ltd.). The concentration x% of LC was adjusted to 0.03%, 0.056%, and 0.065% to obtain (7,3), (8,3), and (9,2) SWCNT fractions, respectively. The obtained aqueous dispersions of SWCNTs were concentrated by centrifugal filtration at 3000 $g$ for 45 min (Hettich Mikro 220R centrifuge, Universal 320R 1615 rotor) using centrifugal filters with 100 kDa molecular weight cutoff (Merck Amicon Ultra-4). The remaining SWCNT dispersion was transferred into an aqueous solution of SDS (0.33 w/vol.%) and adjusted to an optical density (OD) of 0.2 cm$^{-1}$ at the $E_{11}$ transition.

**Functionalization of SWCNTs with sp$^3$ Defects.** (6,5) and (7,5) SWCNTs: PFO-BPy-wrapped (6,5) SWCNTs were functionalized in a 80:20 (vol.-%) mixture of toluene/acetonitrile (MeCN) according to a procedure introduced previously by Berger *et al*.[40] To prevent aggregation of SWCNTs, their concentration was always adjusted to 0.36 mg·L$^{-1}$ in the final reaction mixture (corresponding to an OD of 0.2 cm$^{-1}$). Initially, a solution of 18-crown-6 in toluene was added to a dispersion of SWCNTs such that the final concentration of 18-crown-6 was 7.6 mmol·L$^{-1}$ after addition of the remaining reactants. To achieve a ratio of 80:20 vol.-% of toluene/MeCN, the respective volume of MeCN was then added to the reaction mixture. Appropriate amounts of a stock solution (5 g·L$^{-1}$) of 4-nitrobenzenediazonium tetrafluoroborate (DzNO$_2$) in MeCN were added to achieve concentrations between 20 and 1000 mg·L$^{-1}$. After storage in the dark at room temperature for 16 h, the reaction mixture was filtered through a PTFE membrane filter (Merck Millipore JVWP, pore size 0.1 µm). The filter cake was washed with MeCN (10 mL) and toluene (5 mL) to remove remaining diazonium salt and excess polymer. Redispersion of the sp³-functionalized SWCNTs was performed by bath sonication in fresh toluene. For PFO-wrapped (7,5) SWCNTs, the preparation and workup of the reaction mixture were similar, however, a fixed concentration of 2000 mg·L$^{-1}$ of DzNO$_2$ was used and reaction times were varied between 10 – 200 h to achieve different degrees of functionalization. (7,3), (8,3), and (9,2) SWCNTs: Aqueous dispersions of SWCNTs in 0.33 w/vol.-% SDS were combined with the appropriate volume of an aqueous solution of DzNO$_2$ such that final DzNO$_2$ concentrations between 5·10$^{-5}$ and 5·10$^{-3}$ mg·mL$^{-1}$ were achieved depending on the reactivity of the SWCNT species. The reaction mixture was stored at room temperature in the dark for at least 4 days to ensure completion of the functionalization reaction.

**Characterization.** Baseline-corrected UV-Vis-nIR absorption spectra were collected with a Varian Cary 6000i UV-Vis-nIR absorption spectrometer. All acquired spectra were corrected with a scattering background $S(\lambda) = S_0 e^{-b\lambda}$ according to a literature procedure.[65,66] Raman spectroscopy was carried out with a Renishaw inVia Reflex confocal Raman microscope in backscattering configuration equipped with a ×50 long-working distance objective (Olympus, N.A. 0.5). Samples were prepared by drop-casting the dispersion on glass substrates (Schott AF32eco), followed by thorough washing with tetrahydrofuran and 2-propanol for dispersions of polymer-wrapped SWCNTs or with deionized water in the case of aqueous SWCNT dispersions. Raman spectra were recorded using three different lasers operating at wavelengths of 532 nm (2.33 eV), 633 nm (1.96 eV), and 785 nm (1.58 eV). Over 3000 individual Raman spectra were collected from an area of 60×60 µm², averaged and baseline-corrected for each sample. Raman modes of SWCNTs were analyzed using Lorentzian fits.[23] Photoluminescence excitation-emission maps were recorded with a Fluorolog-3 spectrometer (Horiba Jobin-Yvon) equipped with a xenon arc-discharge lamp (450 W), an excitation double monochromator and a liquid nitrogen cooled linear InGaAs array (Symphony II).

**Photoluminescence Spectroscopy.** For photoluminescence measurements, the output of a picosecond-pulsed supercontinuum laser (NKT Photonics SuperK Extreme) was wavelength-filtered corresponding to the SWCNT $E_{22}$ transition and focused on the nanotube dispersion inside a quartz glass cuvette using a nIR-optimized ×50 objective (Olympus, N.A. 0.65). Emitted light was collected through the same objective and PL spectra were recorded using a grating spectrograph (Princeton Instruments Acton SpectraPro SP2358, grating 150 lines/mm) in combination with a liquid nitrogen cooled InGaAs line camera (Princeton Instruments OMA V:1024 1.7 LN).

**Photoluminescence Quantum Yield Measurements.** Photoluminescence quantum yields (PLQYs) of SWCNT dispersions were determined by an absolute method as described previously.[35] A quartz glass cuvette was filled with 1 mL of SWCNT dispersion (OD = 0.2 cm$^{-1}$ at $E_{11}$) and placed inside an integrating sphere (LabSphere with Spectralon coating). The light exiting the integrating sphere was guided to the spectrometer and detector (see above) with an optical fiber. The PLQY ($\eta$) of each sample was determined as the fraction of emitted ($N_{em}$) to absorbed photons ($N_{abs}$) according to:

$$\eta = \frac{N_{em}}{N_{abs}} \qquad (3)$$

To obtain a value proportional to $N_{em}$, the recorded signal for a solvent reference and the SWCNT dispersion were each multiplied with the wavelength, integrated, and subtracted. An identical analysis was performed with the attenuated laser intensity at the $E_{22}$ transition for the SWCNT sample and a solvent reference to calculate $N_{abs}$ and subsequently the PLQY according to Equation (3). The absorption characteristics of optics in the detection path and wavelength-dependent detector efficiency were accounted for by collection of calibration spectra using a broadband light source (SLS201/M, Thorlabs).

**Calculation of sp³ Defect Densities.** The calculation of luminescent sp³ defect density is based on the model of diffusion-limited contact quenching (DLCQ) as reported previously,[34] which assumes competition between radiative decay of mobile excitons *via* $E_{11}$ or $E_{11}*$ emission and non-radiative decay at quenching sites, for example, nanotube ends.[41] In pristine SWCNTs, the $E_{11}$ PLQY can be calculated according to:

$$\eta = \frac{\pi}{2 \cdot n_q^2 \cdot D \cdot \tau_{rad}} \tag{4}$$

Here, $n_q$ denotes the number of innate quenching sites per μm of nanotube, $D$ the diffusion constant and $\tau_{rad}$ the radiative lifetime of mobile excitons in the $E_{11}$ state. The latter two are independent of the nanotube length and the density of emissive defects or non-radiative quenching sites. In this work, we use values of $D = 10.7 \pm 0.4$ cm²·s⁻¹ and $\tau_{rad} = 3.35 \pm 0.41$ ns for all investigated SWCNT species although these were determined in previous experimental studies only for (6,5) SWCNTs.[42,46]

For functionalized SWCNTs, additional traps for excitons in the form of luminescent sp³ defects are introduced. This reduces the probability for the radiative $E_{11}$ decay of excitons which is reflected in the expression for the $E_{11}$ PLQY of functionalized SWCNTs ($\eta^*$):

$$\eta^* = \frac{\pi}{2 \cdot (n_q + n_d)^2 \cdot D \cdot \tau_{rad}} \tag{5}$$

An expression for the density of sp³ defects $n_d$ is obtained by combining Equations (4) and (5):

$$n_d = n_q \left( \sqrt{\frac{\eta}{\eta^*}} - 1 \right) = \sqrt{\frac{\pi}{2 \cdot \eta \cdot D \cdot \tau_{rad}}} \left( \sqrt{\frac{\eta}{\eta^*}} - 1 \right) \tag{6}$$

From the margins of error for the exciton diffusion constant, the exciton radiative lifetime and the PLQY determination, an overall error of ±15% can assumed for the extracted values of $n_d$.

ACKNOWLEDGEMENTS

This project has received funding from the European Research Council (ERC) under the European Union's Horizon 2020 research and innovation programme (Grant agreement No. 817494, "TRIFECTs"). This study was partially supported by JST CREST (JPMJCR17I5); JSPS KAKENHI (JP22H01911, JP22H05468, JP20H02573, JP21H05017, JP23H00259).

# Supporting Information

# Unified Quantification of Quantum Defects in Small-Diameter Single-Walled Carbon Nanotubes by Raman Spectroscopy

*Finn L. Sebastian[1], Felicitas Becker[1], Yohei Yomogida[2], Yuuya Hosokawa[2], Simon Settele[1], Sebastian Lindenthal[1], Kazuhiro Yanagi[2], Jana Zaumseil[1]\**

[1]Institute for Physical Chemistry, Universität Heidelberg, D-69120 Heidelberg, Germany

[2]Department of Physics, Tokyo Metropolitan University, Tokyo 192-0397, Japan

Corresponding Author
*E-mail: zaumseil@uni-heidelberg.de

# Figures and Tables

**UV-Vis-nIR absorption spectra**

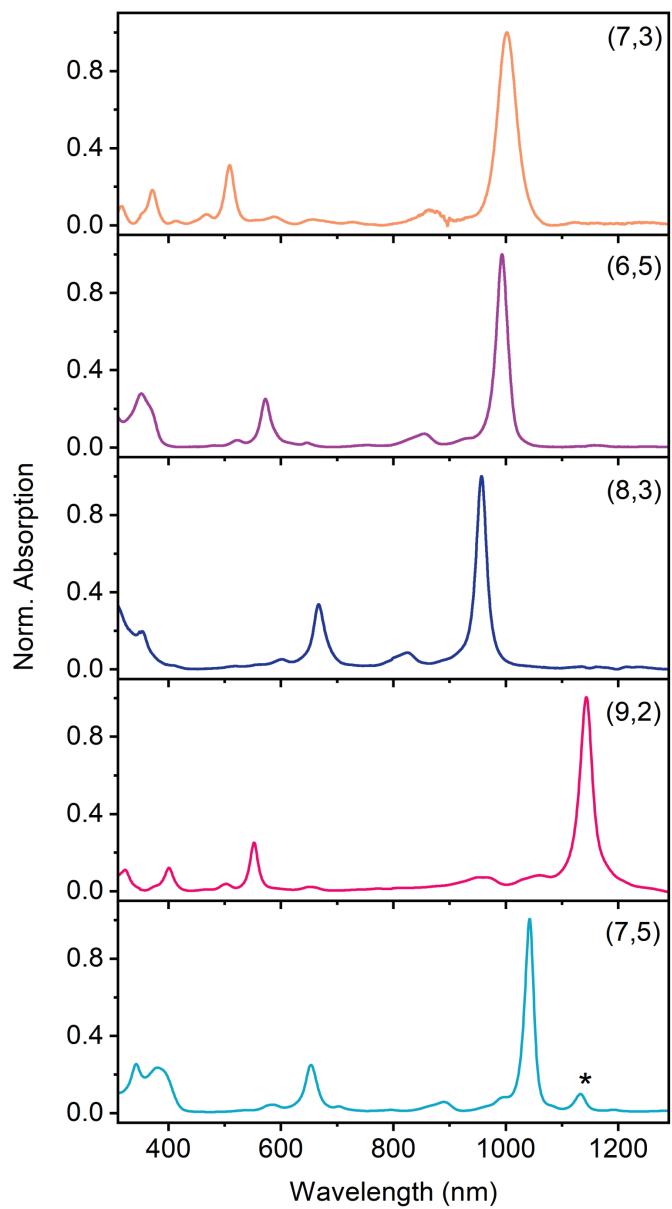

**Figure S1.** (a) Normalized UV-Vis-nIR absorption spectra of pristine (7,3), (6,5), (8,3), (9,2), and (7,5) SWCNTs (residual (7,6) SWCNTs are marked with an asterisk).

**Photoluminescence excitation-emission maps**

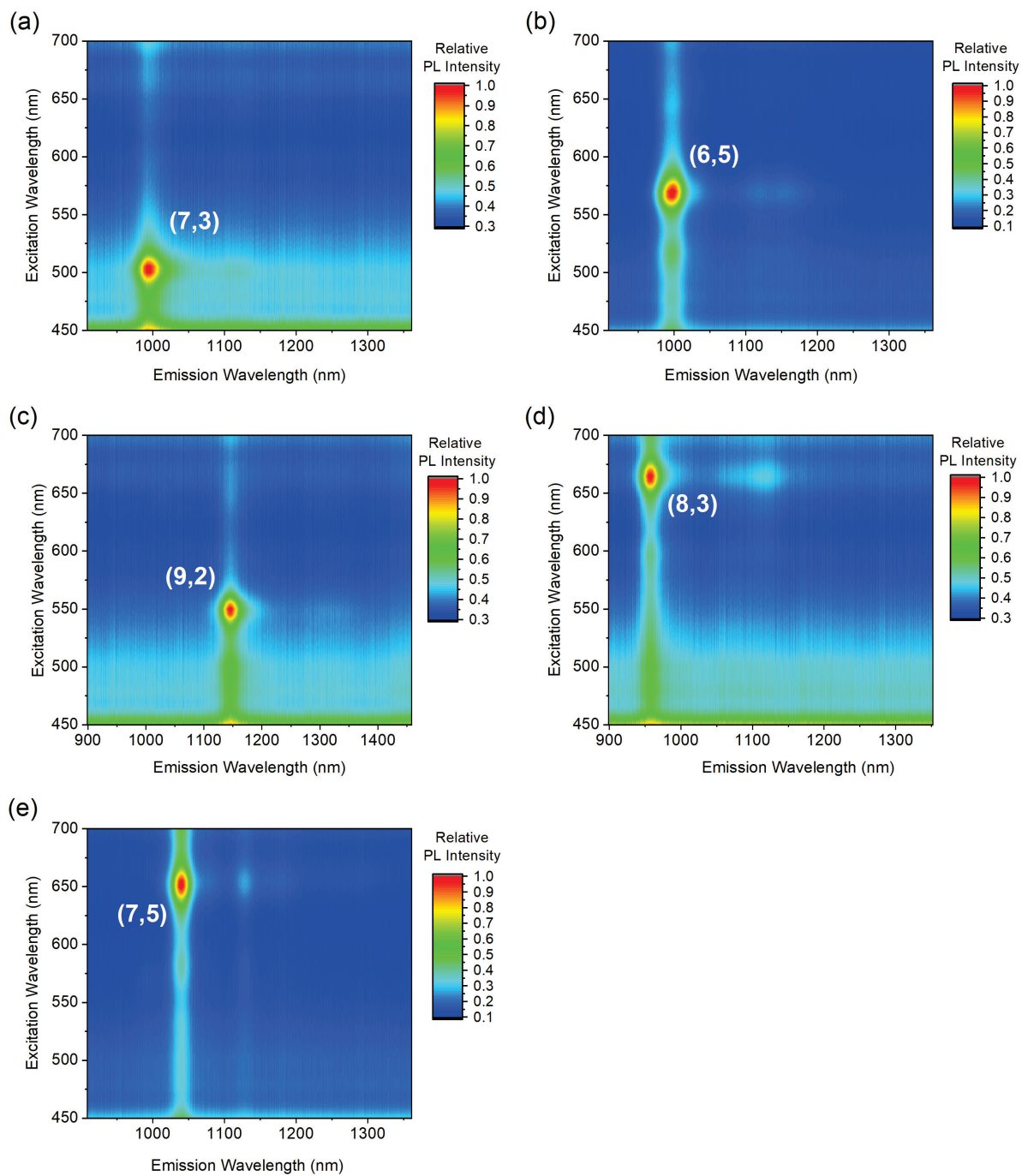

**Figure S2.** Photoluminescence excitation-emission maps of pristine (a) (7,3), (b) (6,5), (c) (8,3), (d) (9,2), and (e) (7,5) SWCNTs.

**Raman spectra of SWCNTs (RBM region) at different excitation wavelengths**

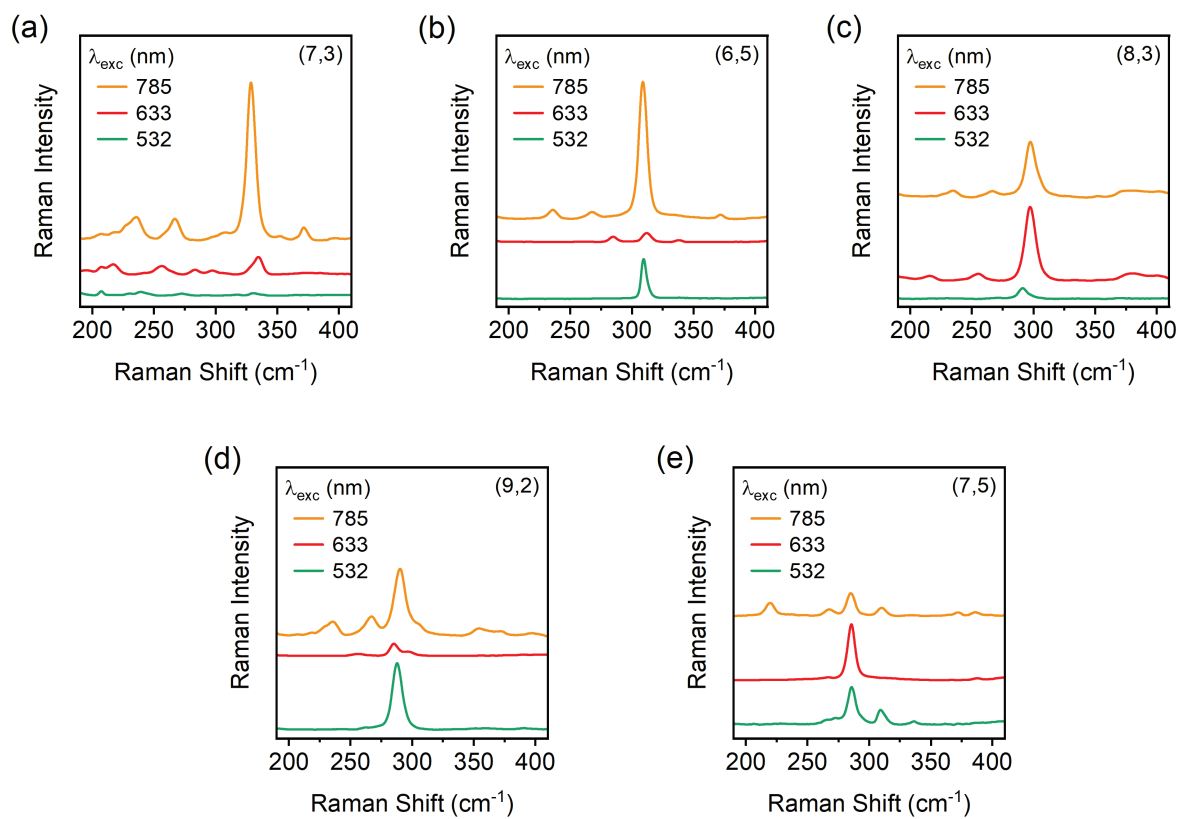

**Figure S3.** Averaged Raman spectra (from >2000 individual spots for each spectrum) of (a) (7,3), (b) (6,5), (c) (8,3), (d) (9,2), and (e) (7,5) SWCNTs in the RBM region, acquired at excitation wavelengths of 532 nm, 633 nm, and 785 nm.

**Diameter dependence of Raman G⁺ and G⁻ mode wavenumber**

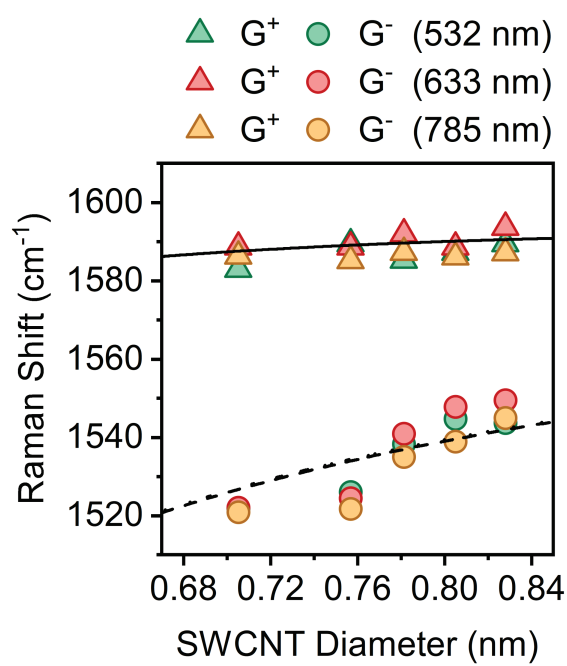

**Figure S4.** Wavenumbers of Raman G⁺ and G⁻ modes obtained from Lorentzian fits to Raman spectra of pristine (7,3), (6,5), (8,3), (9,2), and (7,5) SWCNTs at excitation laser wavelengths of 532 nm, 633 nm, and 785 nm. The solid and dashed lines represent empirical expressions for the diameter dependence of G⁺ and G⁻ modes as reported by Telg *et al.*[1]

## Photoluminescence spectra and PLQYs

### (7,3) SWCNTs

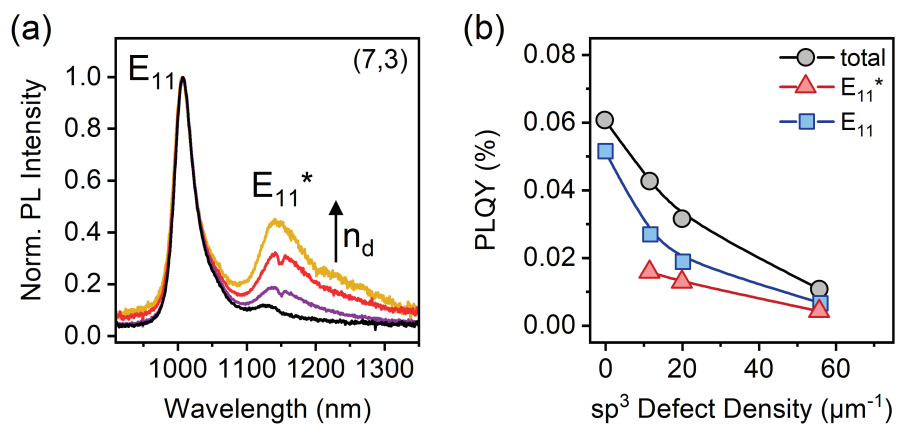

**Figure S5.** (a) Normalized PL spectra under resonant excitation at 509 nm and (b) PLQYs of pristine and sp$^3$-functionalized (7,3) SWCNTs dispersed by SDS in water. Solid lines are guides to the eye.

### (6,5) SWCNTs

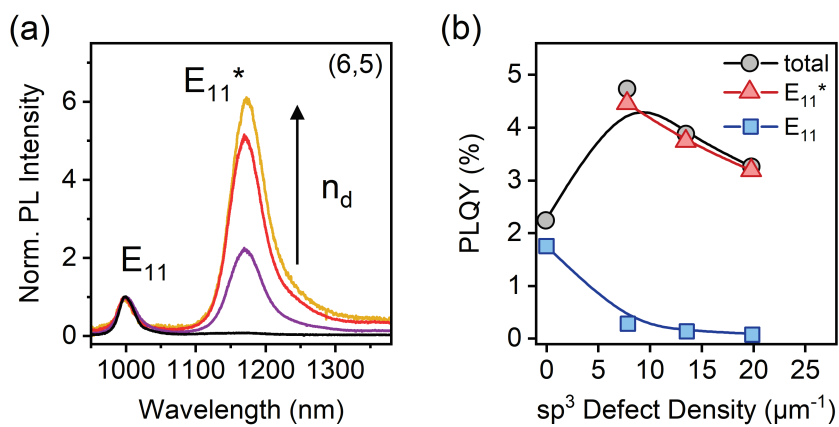

**Figure S6.** (a) Normalized PL spectra under resonant excitation at 575 nm and (b) PLQYs of pristine and sp$^3$-functionalized (6,5) SWCNTs dispersed by PFO-BPy in toluene. Solid lines are guides to the eye.

**(8,3) SWCNTs**

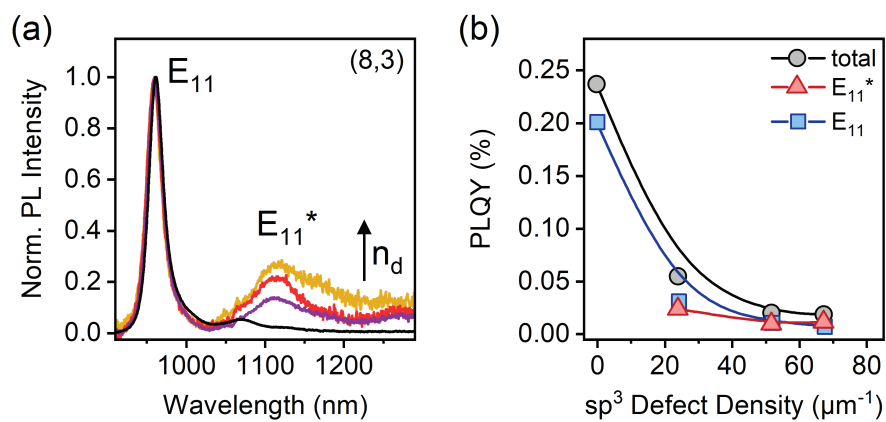

**Figure S7.** (a) Normalized PL spectra under resonant excitation at 668 nm and (b) PLQYs of pristine and sp$^3$-functionalized (8,3) SWCNTs dispersed by SDS in water. Solid lines are guides to the eye.

**(9,2) SWCNTs**

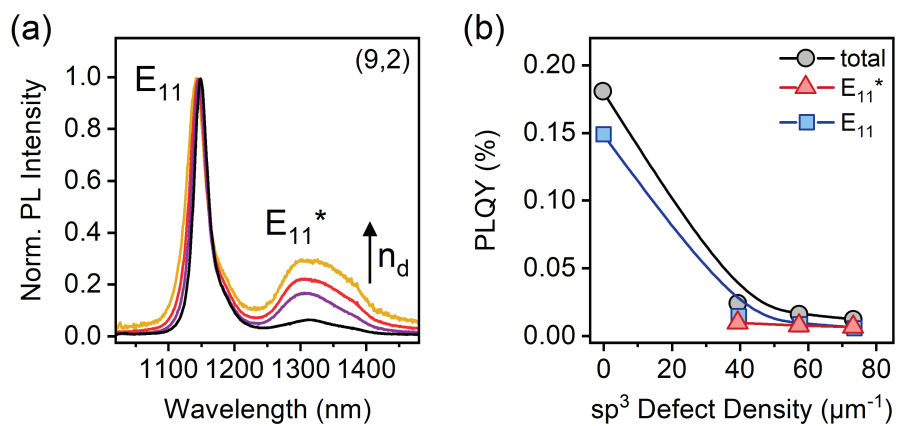

**Figure S8.** (a) Normalized PL spectra under resonant excitation at 553 nm and (b) PLQYs of pristine and sp$^3$-functionalized (9,2) SWCNTs dispersed by SDS in water. Solid lines are guides to the eye.

**(7,5) SWCNTs**

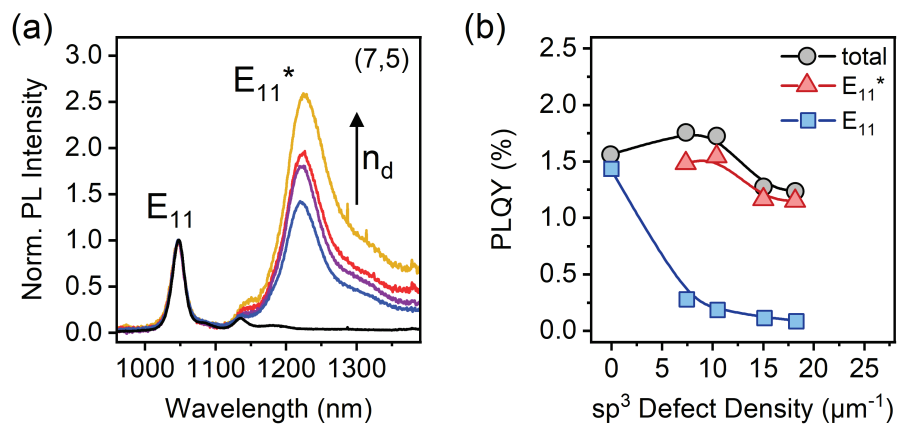

**Figure S9.** (a) Normalized PL spectra under resonant excitation at 652 nm and (b) PLQYs of pristine and sp$^3$-functionalized (7,5) SWCNTs dispersed by PFO in toluene. Solid lines are guides to the eye.

**Raman spectra of SWCNTs at different excitation wavelengths**

**(6,5) SWCNTs**

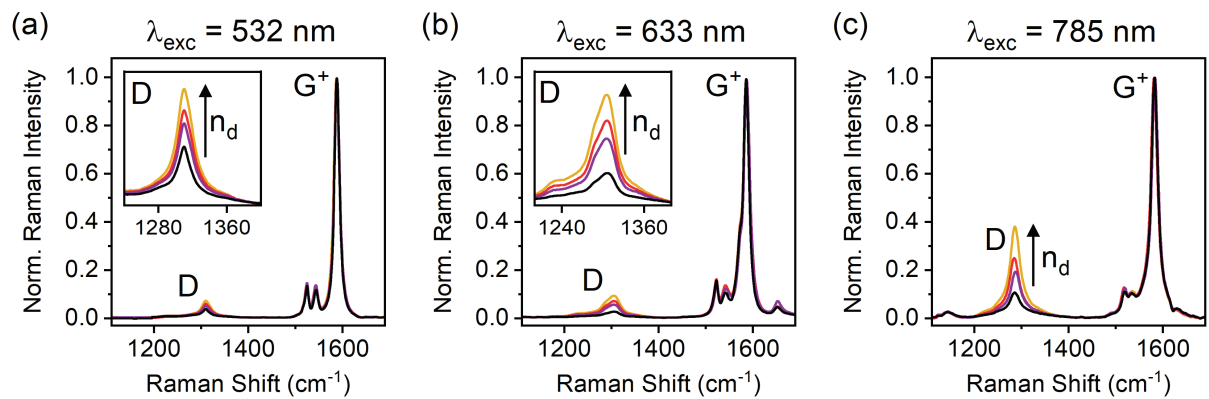

**Figure S10.** Normalized Raman spectra of pristine (black) and sp³-functionalized (6,5) SWCNTs (DzNO2) at different Raman excitation wavelengths of 532 nm (a), 633 nm (b), and 785 nm (c).

**(8,3) SWCNTs**

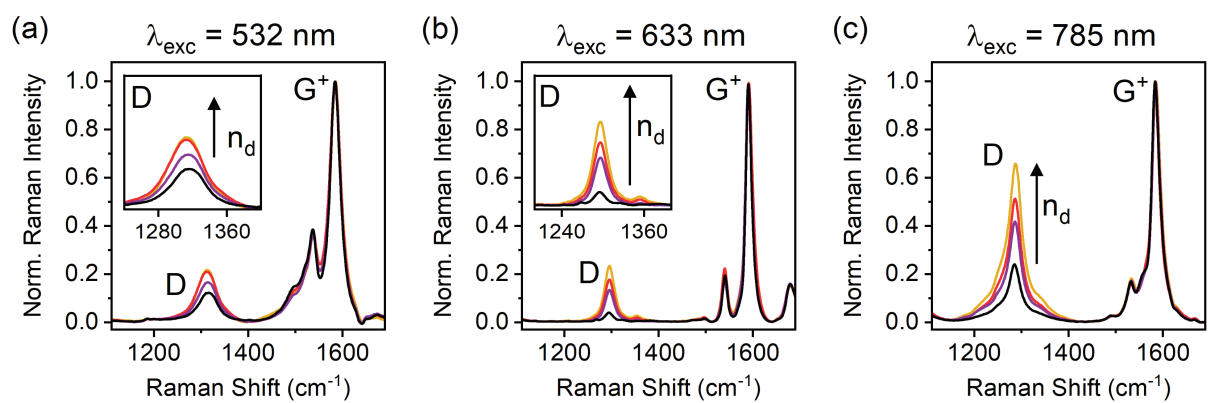

**Figure S11.** Normalized Raman spectra of pristine (black) and sp³-functionalized (8,3) SWCNTs (DzNO2) at different Raman excitation wavelengths of 532 nm (a), 633 nm (b), and 785 nm (c).

**(7,5) SWCNTs**

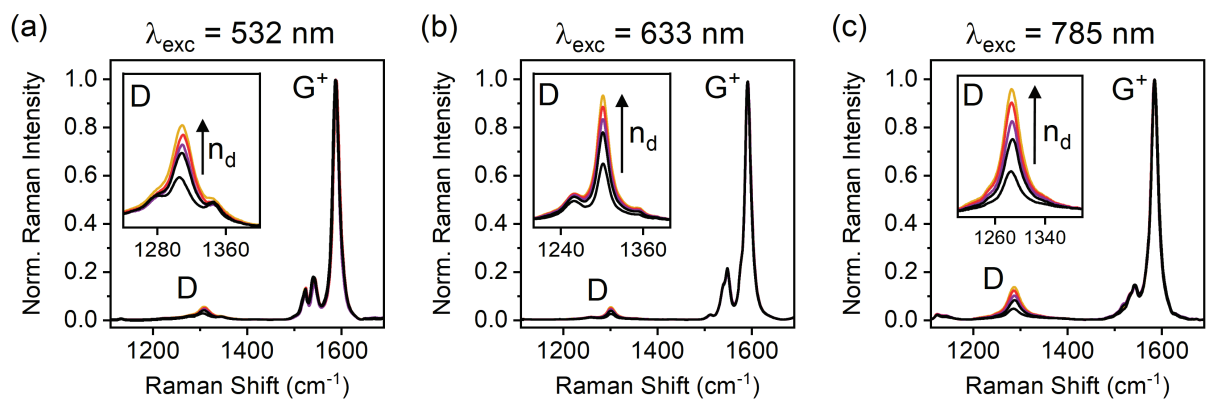

**Figure S12.** Normalized Raman spectra of pristine (black) and sp$^3$-functionalized (7,5) SWCNTs (DzNO$_2$) at different Raman excitation wavelengths of 532 nm (a), 633 nm (b), and 785 nm (c).

**Linear fits of Δ(D/G⁺) *vs* n_d at different excitation wavelengths**

**(7,3) SWCNTs**

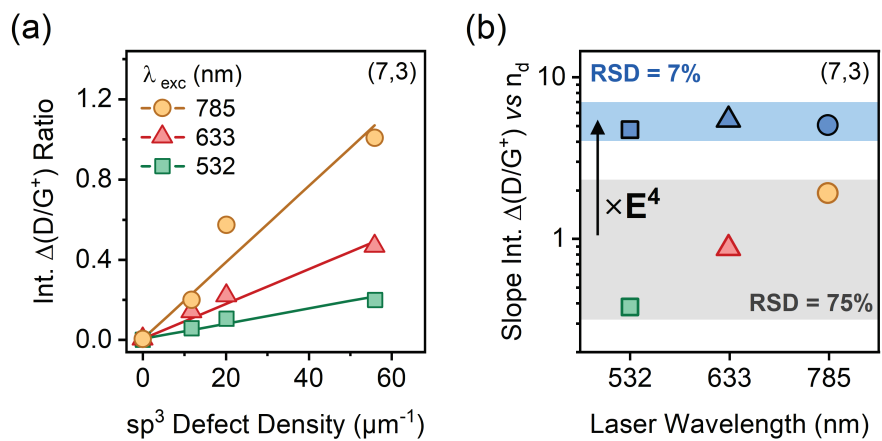

**Figure S13.** (a) Linear fits of integrated Δ(D/G⁺) Raman mode ratio *vs* defect density from PLQY measurements for sp³-functionalized (7,3) SWCNTs at excitation wavelengths of 532 nm ($R^2 = 0.98$), 633 nm ($R^2 = 0.99$), and 785 nm ($R^2 = 0.98$). (b) Slopes of linear fits in (a) for different excitation wavelengths (*i.e.*, excitation energy) and their normalization with excitation energy $E_L^4$ including relative standard deviation (RSD) in %.

**(6,5) SWCNTs**

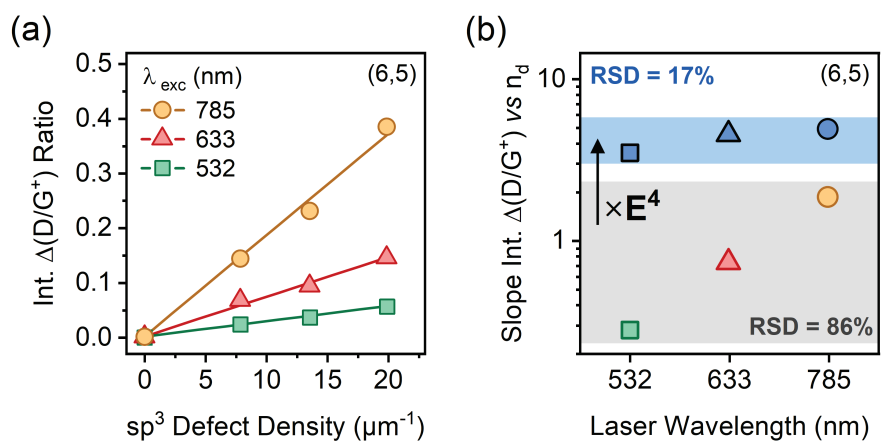

**Figure S14.** (a) Linear fits of integrated Δ(D/G⁺) Raman mode ratio *vs* defect density from PLQY measurements for sp³-functionalized (6,5) SWCNTs at excitation wavelengths of 532 nm ($R^2 = 0.99$), 633 nm ($R^2 = 0.99$), and 785 nm ($R^2 = 0.99$). (b) Slopes of linear fits in (a) for different excitation wavelengths (*i.e.*, excitation energy) and their normalization with excitation energy $E_L^4$ including relative standard deviation (RSD) in %.

**(7,5) SWCNTs**

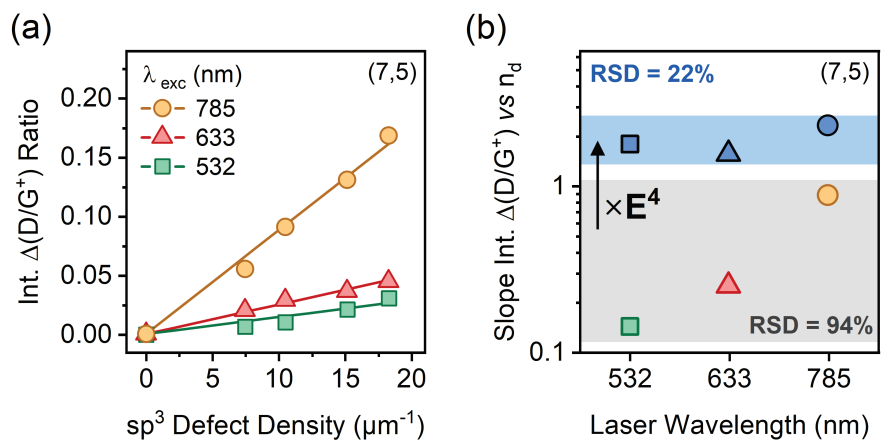

**Figure S15.** (a) Linear fits of integrated Δ(D/G⁺) Raman mode ratio *vs* defect density from PLQY measurements for sp³-functionalized (7,5) SWCNTs at excitation wavelengths of 532 nm ($R^2 = 0.98$), 633 nm ($R^2 = 0.99$), and 785 nm ($R^2 = 0.99$). (b) Slopes of linear fits in (a) for different excitation wavelengths (*i.e.*, excitation energy) and their normalization with excitation energy $E_L^4$ including relative standard deviation (RSD) in %.

**Normalization to the number of carbon atoms per micrometer**

**Table S1.** Number of carbon atoms and SWCNT lattice area per micrometer of carbon nanotube for the investigated chiralities.

| Chirality | (7,3) | (6,5) | (8,3) | (9,2) | (7,5) |
|---|---|---|---|---|---|
| Number of carbon atoms ($\mu m^{-1}$) | 83441 | 89554 | 92458 | 95276 | 98012 |
| Area ($m^2$) | $2.23\times10^{-15}$ | $2.39\times10^{-15}$ | $2.45\times10^{-15}$ | $2.54\times10^{-15}$ | $2.61\times10^{-15}$ |

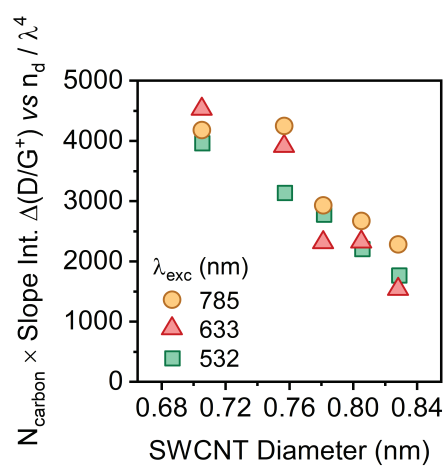

**Figure S16.** Slope of the integrated Raman $\Delta(D/G^+)$ ratio *vs* calculated defect density $n_d$ multiplied with the number of carbon atoms per micrometer of the respective SWCNT chirality.

**Correlation with chiral angle and mod family**

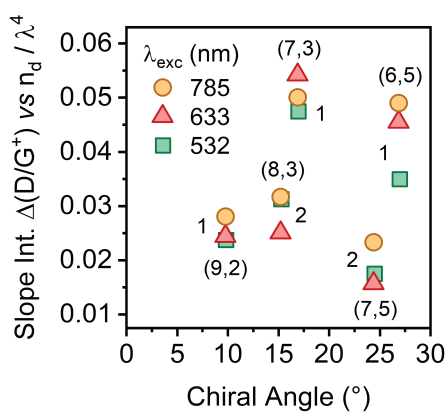

**Figure S17.** Slope of the integrated Raman $\Delta(D/G^+)$ ratio normalized to the excitation wavelength $\lambda_L^{-4}$ *vs* chiral angle $\theta$. (7,3), (6,5), and (9,2) SWCNTs belong to the mod 1 family, whereas (8,3) and (7,5) SWCNTs belong to the mod 2 family.

# Comparison of defect density in graphene and SWCNTs

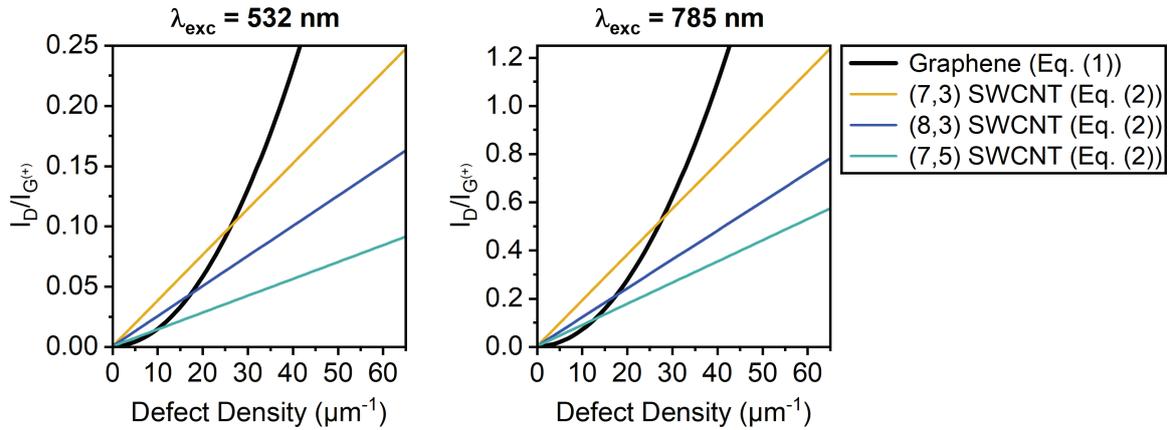

**Figure S18.** (a) Raman $I_D/I_{G^{(+)}}$ ratio *vs* calculated defect density $n_d$ ($\mu m^{-1}$) at excitation wavelengths of 532 nm and 785 nm for graphene and different SWCNT species according to equations (1) and (2) in the main text, respectively.

**Table S2.** Raman $I_D/I_{G^{(+)}}$ ratios at specific calculated defect densities $n_d$ ($\mu m^{-1}$) for all investigated SWCNT chiralities and graphene according to equations (1) and (2) in the main text (Raman excitation wavelength 532 nm).

| $n_d$ ($\mu m^{-1}$) | $I_D/I_{G^{(+)}}$ ($n_d$ ($\mu m^{-1}$)) | | | | | |
|---|---|---|---|---|---|---|
| | (7,3) | (6,5) | (8,3) | (9,2) | (7,5) | Graphene |
| 5  | 0.020 | 0.017 | 0.011 | 0.010 | 0.007 | 0.004 |
| 10 | 0.040 | 0.033 | 0.023 | 0.020 | 0.015 | 0.014 |
| 20 | 0.080 | 0.067 | 0.046 | 0.040 | 0.030 | 0.058 |
| 50 | 0.200 | 0.167 | 0.114 | 0.100 | 0.074 | 0.360 |

## IFM Raman spectra

**(7,3) SWCNTs**

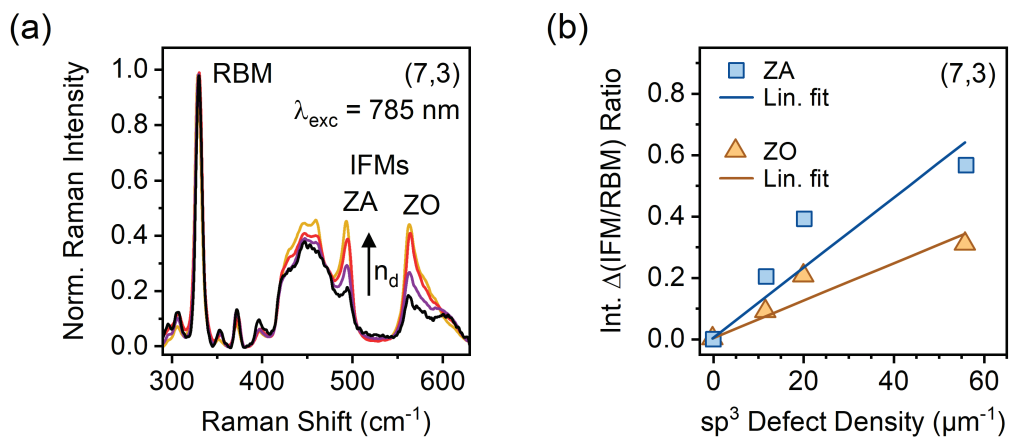

**Figure S19.** (a) Raman spectra of pristine and $sp^3$-functionalized (7,3) SWCNTs in the region of intermediate frequencies acquired at 785 nm excitation. (b) Linear fits of Δ(IFM/RBM) *vs* calculated $sp^3$ defect density $n_d$ for both intermediate frequency modes I and II as marked in (a), $R^2$ (ZA) = 0.96, $R^2$ (ZO) = 0.97.

**(8,3) SWCNTs**

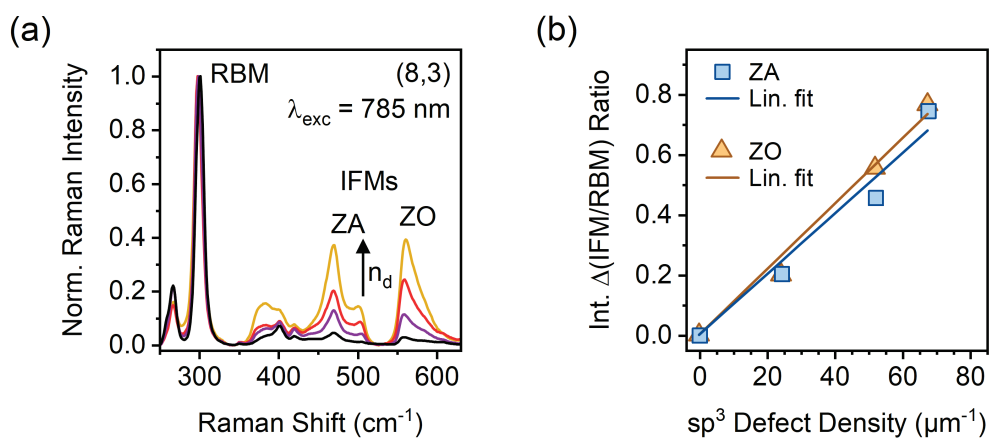

**Figure S20.** (a) Raman spectra of pristine and $sp^3$-functionalized (8,3) SWCNTs in the region of intermediate frequencies acquired at 785 nm excitation. (b) Linear fits of Δ(IFM/RBM) *vs* calculated $sp^3$ defect density $n_d$ for both intermediate frequency modes I and II as marked in (a), $R^2$ (ZA) = 0.99, $R^2$ (ZO) = 0.99.

**(9,2) SWCNTs**

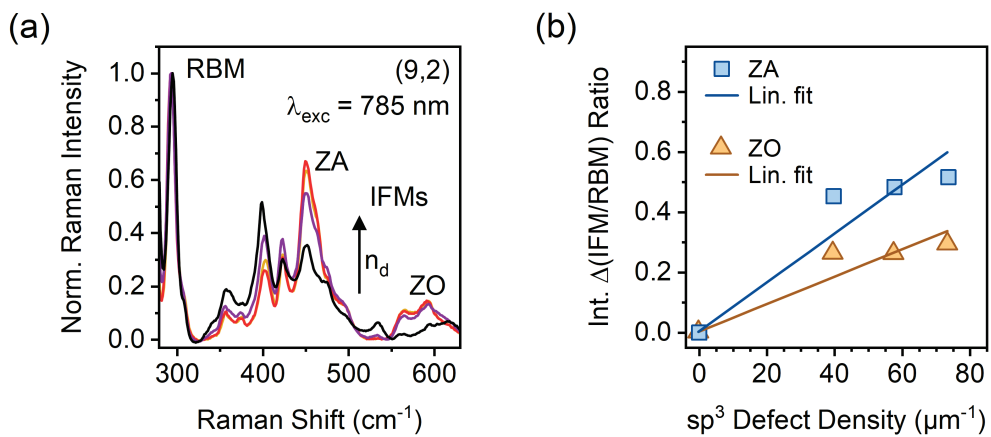

**Figure S21.** (a) Raman spectra of pristine and sp$^3$-functionalized (9,2) SWCNTs in the region of intermediate frequencies acquired at 785 nm excitation. (b) Linear fits of Δ(IFM/RBM) *vs* calculated sp$^3$ defect density n$_d$ for both intermediate frequency modes I and II as marked in (a), $R^2$ (ZA) = 0.98, $R^2$ (ZO) = 0.98.

**(7,5) SWCNTs**

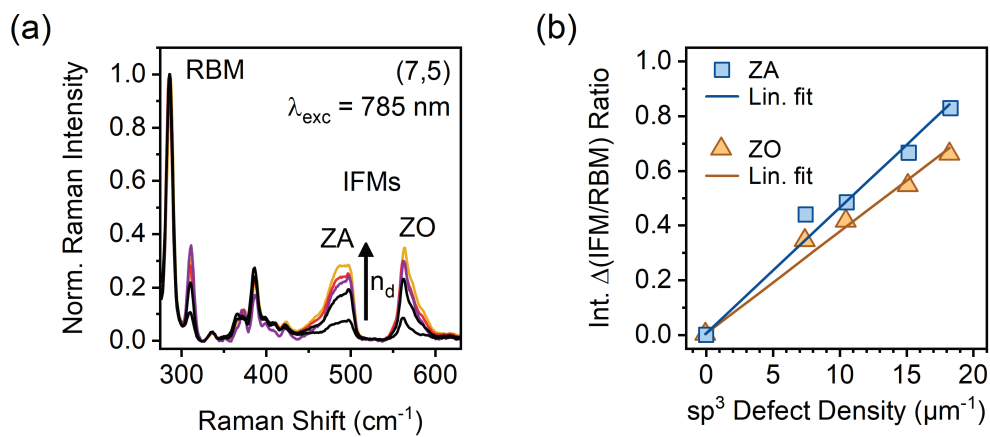

**Figure S22.** (a) Raman spectra of pristine and sp$^3$-functionalized (7,5) SWCNTs in the region of intermediate frequencies acquired at 785 nm excitation. (b) Linear fits of Δ(IFM/RBM) *vs* calculated sp$^3$ defect density n$_d$ for both intermediate frequency modes I and II as marked in (a), $R^2$ (ZA) = 0.99, $R^2$ (ZO) = 0.99.

**Slopes of linear fits Δ(IFM/RBM) *vs* $n_d$**

**Table S3.** Slope (standard deviation) of Δ(IFM/RBM) ratio *vs* calculated sp$^3$ defect density $n_d$ linear fit for different SWCNT chiralities from Raman spectra collected at an excitation wavelength of 785 nm.

| Chirality | (7,3) | (6,5) | (8,3) | (9,2) | (7,5) |
|---|---|---|---|---|---|
| **Slope Δ(ZA/RBM) *vs* $n_d$** | 0.001 (0.002) | 0.038 (0.002) | 0.010 (0.001) | 0.008 (0.001) | 0.046 (0.002) |
| **Slope Δ(ZO/RBM) *vs* $n_d$** | 0.006 (0.001) | 0.021 (0.002) | 0.011 (0.001) | 0.005 (0.001) | 0.037 (0.002) |